\documentstyle[11pt]{article}
\title{\vspace{-1.in} \hfill {\small\rm UTPT-95-10} \\~\\~\\
Chiral Symmetry Breaking via\\ Multi-fermion Green Functions}
\author{ B. Holdom\thanks{Electronic address:holdom$@$utcc.utoronto.ca}
$\;$  and G. Triantaphyllou\thanks{Electronic address:george$@$medb.physics.
utoronto.ca}\\Department of Physics, University of Toronto \\
Toronto, ON M5S 1A7, Canada}
\begin{document}
\renewcommand{\thesection}{\Roman{section}}
\setlength{\baselineskip}{24pt}
\maketitle
\begin{abstract}

Previous results on fermion chirality-flipping four-point functions are
extended to $SU(N)$ gauge theories. The problem is purely
non-perturbative, and it is approached by truncating the
Schwinger-Dyson hierarchy.  The large-$N$ limit also
simplifies the problem substantially. The resulting equation is solved
numerically by relaxation techniques and an estimate of the critical
coupling and momentum behavior
is obtained.  We also comment on the behavior of
chirality-flipping $2n$-point functions for general $n$.
\vspace{2.in}
\end{abstract}
\setcounter{page}{0}
\pagebreak

\section{Introduction}
In this paper we  study the dynamical generation of momentum
dependent fermion
four-point functions within the framework of non-abelian gauge
theories.  These will be purely non-perturbative quantities associated with
breakdown of chiral symmetries, and would for example imply the existence
of the corresponding four-fermion condensates.  Our analysis is based on the
Schwinger-Dyson (SD) formalism.  The resulting  equations are analytically
intractable and thus we seek numerical solutions.

We have two main goals in this study.  One is to estimate the
critical value of the gauge coupling
necessary for the formation of the four-point
functions.  We would like to study this in a case
where four-point functions develop on scales higher than the scale
of mass formation.   This would make it consistent to treat the
four-point function problem in isolation, independently of the mass
generation problem.  This is not the case for QCD, but even in that
case it is useful to at least consider the existence of the effects we
study here.

In extensions of the standard model the hierarchical
symmetry breaking pattern we envisage can be a natural consequence of
the fact that a dynamical four-point function may break fewer continuous
symmetries than a dynamical mass. In the case that some of
these symmetries are gauged, then these additional interactions could
resist the formation of mass.
If so, dynamical masses may form on a scale lower than the scale of the
four-point functions, or not form at all.  With these possibilities in mind,
 we shall simply ignore fermion mass in the present work.

 A multiflavor case will be discussed elsewhere \cite{model}
 in which it is possible to add abelian gauge interactions such as to
 make the theory chiral, so that any mass will break some gauge symmetry.
 In that theory some four-point functions do not break any gauge symmetry;
 this provides a dynamical reason for why four-point functions form and masses
 do not. There are also four-point functions for which the effects of the extra
 $U(1)$'s essentially cancel out, leaving only the attractive non-abelian
interactions.
 The study of such four-point functions would be very similar to the study of
 four-point functions in our simplified single-flavor case, where we assume
 the mass vanishes.

Our other goal is to  extract the momentum
dependence of the dynamical fermion four-point function.  For example if such
an object is to play a role in generating a fermion or a gauge-boson
mass, then two or four lines  of
the four-point function  may be closed off into a loop or into another
four-point function.  It proves then interesting to study
the details of the
momentum dependence, such as the relative size of the four-point
function when different pairs of momenta are large.
The factorization hypothesis, according to which
 the four-point function can be treated as a product of two
two-point functions, is of no use here
since we neglect two-fermion condensates.

In the present study we  use a one gauge-boson exchange approximation.
Since the gauge boson can attach to any pair of the four legs, the SD
equation sums up a much more complicated set of diagrams than the set of
ladder graphs.
The present   work constitutes a clear progress with respect to our previous
study \cite{BG}, since, apart from considering a general
non-abelian group, it includes a treatment of non-linearities,
it does not neglect terms proportional to external momenta while
at the same time exploring the full available momentum space, and
in a certain limit allows us to comment on the
behavior of certain $2n$-point functions.

In the next section we
consider the SD equation for the fermion chirality-flipping
four-point
function in a non-abelian theory, and then compute the form it
takes in the large-$N$ limit of an $SU(N)$ theory in section 3.
In section 4 we discuss our numerical treatment and results, and
in section 5 we generalize our results to $2n$-point functions.

\section{The equations in the non-abelian case}
We focus our
attention on  fermion operators which are purely chirality changing
 of the form $\bar{\psi}_{L}\Gamma\psi_{R}
\bar{\psi}_{L}\Gamma^{\prime}\psi_{R}$ + {\rm h.c.}, where $\psi_{L} \equiv
\frac{(1-\gamma_{5})}{2}\psi$ and $\psi_{R} \equiv
\frac{(1+\gamma_{5})}{2}\psi$.  We constrain our study to just one
fermion flavor in a representation of a gauged simple Lie group.
The four independent operators
which have this property and respect parity are\footnote{We choose to work in
Euclidean space, so there is no difference between upper and lower Lorentz
indices.}
\begin{eqnarray}
\frac{1}{2} (\bar{\psi}\psi \bar{\psi}\psi
 + \bar{\psi}\gamma^{5}\psi \bar{\psi}\gamma^{5}\psi) &=&
\bar{\psi}_{L}\psi_{R}
 \bar{\psi}_{L}\psi_{R} + {\rm h. c.}
 \nonumber \\
 \bar{\psi}\sigma^{\mu\nu}\psi\bar{\psi}\sigma^{\mu\nu}\psi &=&
 \bar{\psi}_{L}\sigma^{\mu\nu}\psi_{R}\bar{\psi}_{L}
 \sigma^{\mu\nu}\psi_{R} +
 {\rm   h. c.} \nonumber \\
\frac{1}{2} (\bar{\psi}\lambda^{a}\psi \bar{\psi}\lambda^{a}\psi
 + \bar{\psi}\lambda^{a}\gamma^{5}\psi \bar{\psi}\lambda^{a}
 \gamma^{5}\psi) &=&
\bar{\psi}_{L}\lambda^{a}\psi_{R}
 \bar{\psi}_{L}\lambda^{a}\psi_{R} + {\rm h. c.}
 \nonumber \\
 \bar{\psi}\lambda^{a}\sigma^{\mu\nu}\psi\bar{\psi}\lambda^{a}
 \sigma^{\mu\nu}\psi &=&
 \bar{\psi}_{L}\lambda^{a}\sigma^{\mu\nu}\psi_{R}\bar{\psi}_{L}
 \lambda^{a}\sigma^{\mu\nu}\psi_{R} +
 {\rm   h. c.},
 \end{eqnarray}
\noindent where $\lambda^{a}$ are the generators of the non-abelian
Lie algebra.  The four-point functions corresponding to these four operators
receive exclusively
non-perturbative contributions. From now on
we will denote all operators of the form
$\bar{\psi} \Gamma \psi \bar{\psi} \Gamma^{\prime} \psi$,
where $\Gamma, \Gamma^{\prime}$ are matrices with possibly
non-trivial spinor and color structure, by $\Gamma \otimes \Gamma^{\prime}$.

The chiral $U(1)_{A}$ is
anomalous, but its discrete subgroup
$e^{i(n/2) \pi\gamma_{5}}$ where $n$ is an integer is still
present and would have to be broken
before fermion mass could form. As mentioned in the introduction, other effects
in a more realistic theory could resist the breaking of such a
symmetry.  Here we shall simply assume that fermion masses are smaller than the
typical scale of our problem, and can thus be ignored.

Note that we have omitted chirality flipping operators
with derivatives, having structures like
$\partial^{\mu}_{i}\partial^{\nu}_{j}\sigma^{\mu \nu}\otimes ({\bf 1}
\; {\rm or} \;
\partial^{\rho }_{k}\partial^{\tau}_{l}\sigma^{\rho \tau})$ and
$\partial_{i}^{\mu}\sigma^{\mu \nu}\otimes \partial_{j}^{\rho}
\sigma^{\rho \nu}$, where each of the derivatives with indices $i, j, k,
l = 1,...,4$ acts on one of the four
 fermion fields. We will present arguments supporting
their omission  later.

\begin{figure}[p]
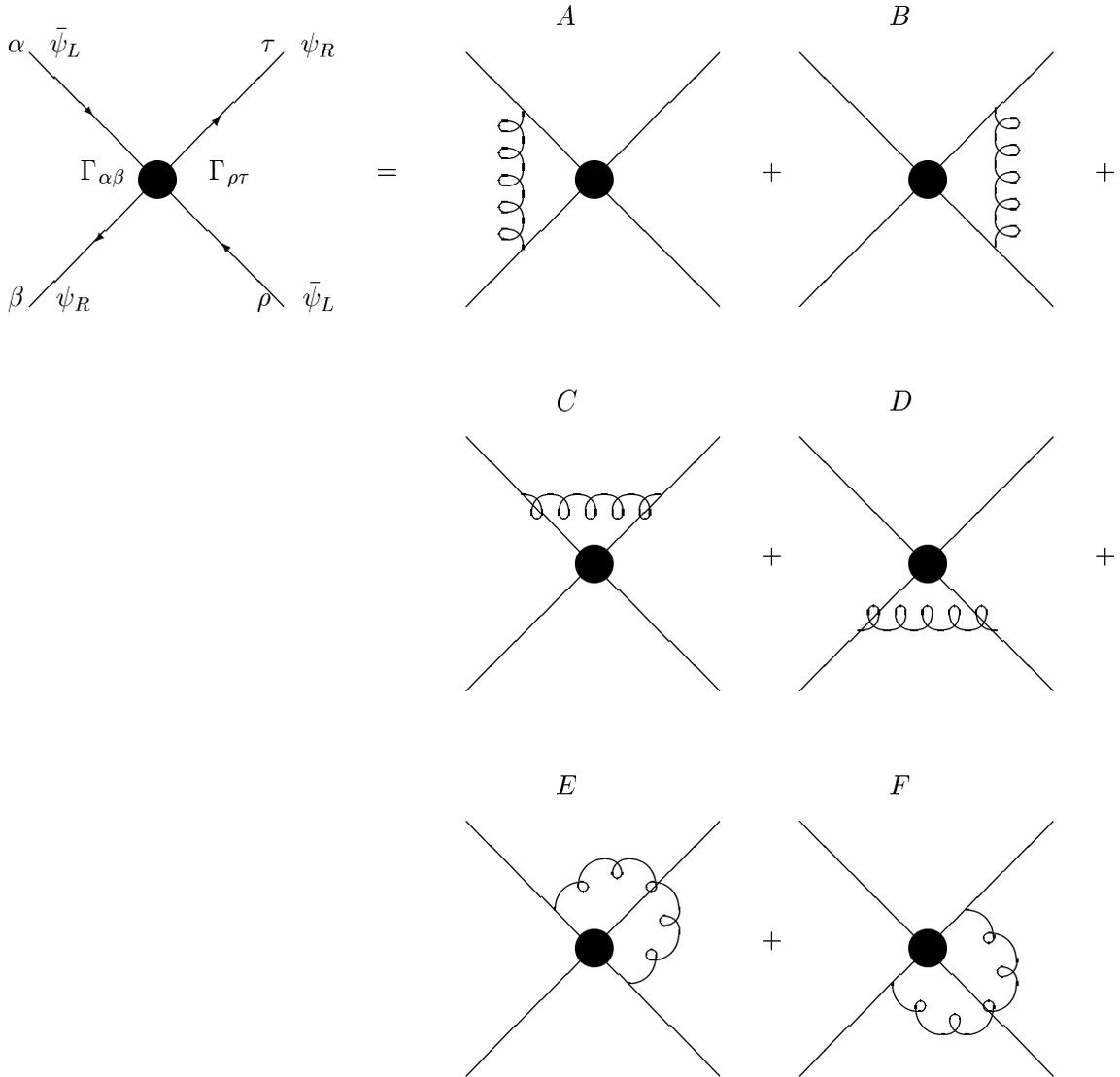

%\vspace{14.cm}
%\special{psfile=SD4.ps hoffset=-100 voffset=-200 angle=0
%                       hscale=100  vscale=100}
\caption{The schematic form of the SD equation.
We have labeled the four fermions
by their spinor indices.
We label the  diagrams
by the capital letters $A,...,F$, and we omit the factor of $1/2$ multiplying
the right-hand side.}
\label{fig:SD4}
\end{figure}

The SD formalism relevant to these four-point functions was discussed in Ref.
\cite{BG} and led to an equation shown diagrammatically in Fig.~\ref{fig:SD4}.
It results from a simple truncation of the SD hierarchy achieved by
approximating the five-point function involving four fermions and a gluon by
the
sum of tree graphs
involving the four-point function. (The equation has been symmetrized to
include diagrams with gluons connecting all possible pairs of fermions, so the
right-hand side is multiplied by a factor of $1/2$.) The analogous procedure
used in the SD equation for the two-point function yields the popular ladder
approximation.  In our case the set of diagrams being resummed has a more
complicated structure than sets of ladder graphs.  In both cases gauge
self-interactions no longer appear in the SD equation.  In the two-point case
it
is then true that there is little difference between abelian and non-abelian
interactions.  On the contrary in the four-point case we show that there is
still a big difference; for non-abelian interactions the space of four point
functions is enlarged and the mixing between the four-point functions is
completely different.

We are considering a four-point function associated with the Green function
\linebreak  $\langle 0|T\{\bar{\psi}_{\alpha}\psi_{\beta}\bar{\psi}_{\rho}
\psi_{\tau}\}|0
\rangle$.
In momentum space, the
four-point function involving scalar functions of momentum only and
receiving exclusively non-perturbative contributions is
\begin{eqnarray}
{{\cal
O}}_{\alpha \beta \rho \tau} & = &
{{\cal
O}}_{S+P}({I}_{\alpha \beta}\otimes {I}_{\rho \tau}+
{\gamma }_{\alpha \beta}^{5}\otimes {\gamma
}_{\rho \tau}^{5})+{{\cal O}}_{T}{\sigma }_{\alpha \beta}^{\mu \nu }
\otimes {\sigma
}_{\rho \tau}^{\mu \nu } \nonumber \\
&& +{{\cal
O}}_{S+P}^{color}(\lambda^{a}{I}_{\alpha \beta}\otimes \lambda^{a}{I}_{
\rho \tau}+
\lambda^{a}{\gamma }_{\alpha \beta}^{5}\otimes \lambda^{a}{\gamma
}_{\rho \tau}^{5})+{{\cal O}}_{T}^{color}\lambda^{a}{\sigma }_{
\alpha \beta}^{\mu \nu }\otimes
\lambda^{a}{\sigma }_{\rho \tau}^{\mu \nu }
\end{eqnarray}
where
the four scalar functions ${\cal O}_{S+P}$,  ${\cal O}_T$,
${\cal O}^{color}_{S+P}$ and ${\cal O}^{color}_T$ depend on
6 variables, which are all the independent
and  Lorentz-invariant combinations of
the external
4-momenta $p_{1}, ..., p_{4}$,
corresponding to the fermions with spinor indices $\alpha,
\beta, \rho, \tau$
respectively. We wish to develop the SD equations for
these scalar functions.

Similarly to our previous work \cite{BG}, we begin by
defining the following
functional operators $\Gamma^{i}$: \begin{eqnarray}
\Gamma^{A}[K] &\equiv& \frac{\alpha}{8\pi^{3}}\int d^{4}k
\frac{K}{(p_{1}-k)^{2}(p_{2}+k)^{2}} \nonumber \\  \Gamma^{B}[K] &\equiv&
\frac{\alpha}{8\pi^{3}}\int d^{4}k \frac{K}{(p_{3}+k)^{2}(p_{4}-k)^{2}}
\nonumber \\  \Gamma^{C}[K] &\equiv& \frac{\alpha}{8\pi^{3}}\int d^{4}k
\frac{K}{(p_{1}-k)^{2}(p_{4}-k)^{2}} \nonumber \\  \Gamma^{D}[K] &\equiv&
\frac{\alpha}{8\pi^{3}}\int d^{4}k \frac{K}{(p_{2}+k)^{2}(p_{3}+k)^{2}}
\nonumber \\  \Gamma^{E}[K] &\equiv& \frac{\alpha}{8\pi^{3}}\int d^{4}k
\frac{K}{(p_{1}-k)^{2}(p_{3}-k)^{2}} \nonumber \\  \Gamma^{F}[K] &\equiv&
\frac{\alpha}{8\pi^{3}}\int d^{4}k \frac{K}{(p_{2}+k)^{2}(p_{4}+k)^{2}},
\end{eqnarray} where $K$ is a function of the loop and external momenta,
with a possibly non-trivial spinor structure, the
letters $A,...,F$ correspond to the diagrams shown in Fig.\ref{fig:SD4}
with the gauge boson having four-momentum $k$,
and $\alpha$ is the
momentum-independent coupling.  We work in the Landau gauge which is
popular in studies of SD equations; the gauge boson propagator reads
$\frac{D^{\mu\nu}}{k^{2}} \equiv \frac{1}{k^{2}}
\left(\delta^{\mu\nu}-\frac{k^{\mu}k^{\nu}}{k^{2}}\right)$.
The choice of this gauge will be further justified later, when we consider
the large-$N$ limit of an $SU(N)$ gauge theory.

By combining the results of the abelian case \cite{BG} with the
study on the color structure given in  Appendix A, and considering
fermions in the fundamental representation of $SU(N)$,  we have
\begin{eqnarray}
{\cal O}_{S+P} & = & \frac{3(N^{2}-1)}{2N}(\overline{\Gamma^{A}}+
\overline{\Gamma^{B}})[{\cal O}_{S+P}]
+ 6(\overline{\Gamma^{C}}+\overline{\Gamma^{D}}- \overline{\Gamma^{E}}
-\overline{\Gamma^{F}})[{\cal O}_{T}^{color}]  \nonumber \\
&& \nonumber \\
{\cal O}_{T} & = & -\frac{N^{2}-1}{2N}(\overline{\Gamma^{A}}+
\overline{\Gamma^{B}})[{\cal O}_{T}] +2(
\overline{\Gamma^{C}}
+\overline{\Gamma^{D}}
+\overline{\Gamma^{E}}
+\overline{\Gamma^{F}})[{\cal O}_{T}^{color}] + \nonumber \\
&& \nonumber \\
& & \frac{1}{2}(
\overline{\Gamma^{C}}
+\overline{\Gamma^{D}}
-\overline{\Gamma^{E}}
-\overline{\Gamma^{F}})[{\cal O}_{S+P}^{color}]  \nonumber \\
&& \nonumber \\
{\cal O}_{S+P}^{color} & = & -\frac{3}{2N}(\overline{\Gamma^{A}}+
\overline{\Gamma^{B}})[{\cal O}_{S+P}^{color}] +\frac{3(N^{2}-1)}{2N^{2}}(
\overline{\Gamma^{C}}
+\overline{\Gamma^{D}}
-\overline{\Gamma^{E}}
-\overline{\Gamma^{F}})[{\cal O}_{T}] + \nonumber \\
&& \nonumber \\
& & 6\left(-\frac{N^{2}+2}{2N}(\overline{\Gamma^{C}}+\overline{\Gamma^{D}})
+\frac{1}{N}(\overline{\Gamma^{E}}+\overline{\Gamma^{F}})\right)[
{\cal O}_{T}^{color}] \nonumber \\
&& \nonumber \\
{\cal O}_{T}^{color} & = &  \left(\frac{1}{2N}(\overline{\Gamma^{A}}+
\overline{\Gamma^{B}})+
\left(-\frac{N^{2}+2}{N}(\overline{\Gamma^{C}}+\overline{\Gamma^{D}}) -
\frac{2}{N}(\overline{\Gamma^{E}}+\overline{\Gamma^{F}})\right)\right)
[{\cal O}_{T}^{color}]
+ \nonumber \\
&& \nonumber \\ && \frac{N^{2}-1}{2N^{2}}
(\overline{\Gamma^{C}}
+\overline{\Gamma^{D}}
+\overline{\Gamma^{E}}
+\overline{\Gamma^{F}})[{\cal O}_{T}]
+ \frac{N^{2}-1}{8N^{2}}
(\overline{\Gamma^{C}}
+\overline{\Gamma^{D}}
-\overline{\Gamma^{E}}
-\overline{\Gamma^{F}})[{\cal O}_{S+P}] + \nonumber \\
&& \nonumber \\
& & \frac{1}{2}
\left(-\frac{N^{2}+2}{2N}(\overline{\Gamma^{C}}+\overline{\Gamma^{D}})+
\frac{1}{N}(\overline{\Gamma^{E}}+\overline{\Gamma^{F}})\right)
[{\cal O}_{S+P}^{color}],
\label{eq:central}
\end{eqnarray}

\noindent
where $\overline{\Gamma^{i}}[{\cal O}] \equiv \Gamma^{i}({\cal O}^{i})$
 and ${\cal O}^{i}$ are the form factors with loop-dependent
 arguments
 corresponding to diagrams $i=A,...,F$ \cite{BG}.

We are faced with a system of four coupled 4-dimensional integral equations
involving functions
of 6 variables. As it stands, the problem is analytically intractable, and
even a numerical solution proves to be beyond our present means.
In the next
section we present the simplification that the large-$N$ limit of $SU(N)$
provides.

\section{The large-$N$ limit}
 By taking the large-$N$ limit, (\ref{eq:central}) reduces to
\footnote{Note that we apply the large-$N$ limit only within the
context of the truncated SD equation, and we do not claim that
the large-$N$ limit in any way justifies the original truncation of the SD
hierarchy.}

\begin{eqnarray}
&&\hspace{-2cm} \left( \begin{array}{c}
{\cal O}_{S+P}
\\ ~\\ {\cal O}_{T} \\~\\ {\cal O}_{S+P}^{color} \\~\\ {\cal O}_{T}^{color}
\end{array} \right) =
N \;\left( \begin{array}{cccc}
\frac{3}{2}(\overline{\Gamma^{A}}+ \overline{\Gamma^{B}})
& 0 & 0 & 0 \\ &&&\\
0 &
-\frac{1}{2}(\overline{\Gamma^{A}}+\overline{\Gamma^{B}})
& 0 & 0 \\ &&&\\
0 & 0 & 0 &
-3(\overline{\Gamma^{C}}+\overline{\Gamma^{D}})
\\ &&&\\
0 & 0 &
-\frac{1}{4}(\overline{\Gamma^{C}}+\overline{\Gamma^{D}})
& -(\overline{\Gamma^{C}}+\overline{\Gamma^{D}})
\end{array} \right)
\left( \begin{array}{c}
\left[ {\cal O}_{S+P}\right] \\~\\
\left[ { \cal O}_{T}\right] \\~\\ \left[ { \cal O}_{S+P}^{color}\right]
\\~\\ \left[ { \cal O}_{T}^{color}\right]
\end{array} \right)
\nonumber \\ &&
\nonumber \\ &&
\end{eqnarray}
The form factors  ${\cal O}_{S+P}$ and ${\cal O}_{T}$ decouple from the others
and from each other.  The ${\cal O}_{S+P}$ entry is positive, which means
that there is
an attractive interaction in this channel necessary for the formation of a
non-zero four-point function. In contrast ${\cal O}_{T}$ is not expected
to be non-zero.
It is also apparent that ${\cal O}_{S+P}^{color}$ can be non-zero
only if ${\cal O}_{T}^{color}$ is somehow non-zero, but that seems
unlikely because of the negative entry coming from   ${\cal O}_{T}^{color}$
itself.
Even if the ${\cal O}_{S+P,T}^{color}$ functions were somehow non-zero,
we would expect for them a critical coupling
larger than the one required for
${\cal O}_{S+P}$.

The large-$N$ limit also allows us to argue for the omission of
contributions coming from  operators proportional to external momenta.
With regards to the operators with structure
$p_{i}^{\mu}\sigma^{\mu \nu}\otimes p_{j}^{\lambda}\sigma^{\nu \lambda}$
and $p_{i}^{\mu}p_{j}^{\nu}\sigma^{\mu \nu}\otimes
p_{k}^{\lambda}p_{l}^{\tau}\sigma^{\lambda \tau}$, we can see that the
contributions they receive from the dominant diagrams
having the same functions in their vertices
(diagrams $A$ and $B$ for
${\bf 1} \otimes {\bf 1}$ color structures, diagrams $C$ and $D$ for
$\lambda^{a} \otimes
\lambda^{a}$ color structures)
are negative, so we do not expect the corresponding four-point functions
to be non-zero. This happens because, in the large-$N$ limit,
tensor insertions
give negative contributions.  Even if
other terms made them non-zero, they would require a larger
gauge coupling than the one needed for $O_{S+P}$.

Interesting also are the operators with structure
${\cal O}_{1}\equiv p^{\mu}_{i}p_{j}^{\nu} \sigma^{\mu \nu} \otimes {\bf 1}$
and ${\cal O}_{2}\equiv {\bf 1} \otimes
p^{\mu}_{i}p_{j}^{\nu} \sigma^{\mu \nu}$ (and trivial color structure),
which can obviously mix with
the form factor ${\cal O}_{S+P}$. If we take the operator ${\cal O}_{1}$
for instance, we see that it  receives a negative contribution from
diagram $A$ and a positive one from diagram $B$. Therefore, it might
 not develop a non-zero value, and if it did, it would require a larger
 value for the gauge coupling than the critical value
 we are seeking. These operators will also be neglected.

We now write down the integral equation for ${\cal O}_{S+P}$.
We go to a reference frame where $\vec{p_{1}}=-\vec{p_{2}}$,
where the form of the kernel of the integral equation indicates that
${\cal O}_{S+P}$ is a function of five variables, i.e.
\begin{eqnarray}
{\cal O}_{S+P}
(p_{1}^{0}, |\vec{p_{1}}|, p_{4}^{0},
|\vec{p_{4}}|, q^{0})
 &=& \frac{N\alpha}{16\pi^{3}}\int d^{4}k
\gamma^{\mu}\frac{1}{k\!\!\!/(
k\!\!\!/ - q\!\!\!/)}\gamma^{\nu}  \nonumber \\ &&
\hspace{-2in} \times \left(\frac{{\cal O}_{S+P}
(k^{0}, |k|, p_{4}^{0}, |\vec{p_{4}}|, q^{0})
D_{1}^{\mu\nu}
}{(k-p_{1})^{2}}+\frac{{\cal O}_{S+P}
(p_{1}^{0}, |\vec{p_{1}}|, k^{0}, |k|, q^{0})
D_{4}^{\mu \nu}}{(k-p_{4})^{2}
}\right)
\end{eqnarray}
\noindent where the two terms on the right-hand
side come from diagrams $A$ and $B$
respectively,  $q \equiv p_{1}+p_{2}$,
$D^{\mu \nu}_{i} \equiv \delta^{\mu \nu}-
\frac{(k-p_{i})^{\mu}(k-p_{i})^{\nu}}{(k-p_{i})^{2}}$, and the
two fermions in the loop have four-momenta $k$ and $k-q$
respectively.

We find that a solution to the above integral
equation takes the factorized form ${\cal O}_{S+P} \sim
\tilde{B}(p^{0}_{1}, |\vec{p_{1}}|, q^{0})\tilde{B
}(p^{0}_{4}, |\vec{p_{4}}|, q^{0})$.  This just reflects the fact
that the loop integral in diagram $A$ does not depend on $p^{0}_{4}$ and
$|\vec{p_{4}}|$ for a given $q^{0}$, and similarly for diagram $B$. By
momentarily setting $p^{0}_{1}=p^{0}_{4} \equiv p^{0}$ and
$|\vec{p_{1}}|=|\vec{p_{4}}| \equiv |p|$ we obtain an equation
for $\tilde{B}$.
\begin{equation}
\tilde{B}(p^{0}, |p|; q^{0}) = \frac{N\alpha}{8\pi^{3}}\int d^{4}k
\tilde{B}(k^{0}, |k|; q^{0})
\gamma^{\mu}\frac{1}{k\!\!\!/(
k\!\!\!/ - q\!\!\!/)}\gamma^{\nu}\frac{D_{1}^{\mu\nu}
}{(k-p_{1})^{2}}
\end{equation}
\noindent We have separated $q^{0}$ from the other two variables
with a semi-colon since
it is an argument of $\tilde{B}$ that is not affected by loop momenta,
and it can thus be treated as if it were
a parameter in the kernel.

Note that for
$q^{0}=0$ this reduces to the linearized
SD equation for the fermion two-point function.
For non-zero $q^{0}$, the problem reduces to the one of a three-point
function, which has been studied before \cite{Aoki}.
After some  Dirac algebra, and omitting terms
with structure $p_{i}^{\mu}p_{j}^{\nu}\sigma^{\mu \nu}\otimes {\bf 1}$
according to our previous discussion,
the equation
becomes
\begin{eqnarray}
\tilde{B}(p^{0}, |p|; q^{0}) &=& \frac{3N\alpha}{4\pi^{2}}\int dk^{0}d|k|
\frac{\tilde{B}
(k^{0}, |k|; q^{0})|k|}{\left((q^{0}-k^{0})^{2}+|k|^{2}\right)2|p|}
\times \nonumber \\
&& \nonumber \\ &&
\ln{\left(\frac{
(|k|+|p|)^{2}+(k^{0}-p^{0})^{2}}
{(|k|-|p|)^{2}+(k^{0}-p^{0})^{2}}
\right)}
\left(1- \frac{k^{0}q^{0}}{k^{2}}\right)
\end{eqnarray}
\noindent
In the following we will change our variables from $p^{0}$, $k^{0}$,
$|p|$, and $|k|$, to $p= \sqrt{p^{0\;2}+|p|^{2}}$, $k=\sqrt{
k^{0\;2}+|k|^{2}}$, $\phi = \arctan{(|p|/p^{0})}$ and
$\tilde{\phi}=\arctan{(|k|/k^{0})}$.
Without loss of generality  we will write $\tilde{B}(p^{0}, |p|; q^{0})
\equiv R(q^{0})B(p, \phi; q^{0})$ with $R(0)\equiv 1$ and
$B(0,0; q^{0})\equiv 1$.

At this stage our equation is linear, and the function $R(q^0)$ remains
undetermined. Clearly some essential physics, containing the non-linearities,
has been omitted by the original truncation. In the two-point function case it
is known that the main effect of the non-linearities can be modeled by
introducing an infrared cutoff, which is determined in a self-consistent manner
by identifying it with the dynamical mass evaluated at the infrared cutoff. We
will follow a similar procedure in the four-point function case by identifying
the infrared cutoff with $\Lambda R(q^0)$, where $\Lambda$ is the infrared
cutoff when $q^0=0$, and then determining $R(q^0)$ in a self-consistent manner.

Non-linear effects will enter through diagrams involving an odd number of
four-point functions, an example of which is provided by the diagram in
Fig.\ref{fig:SD3}. For large gluon momentum this diagram is much more
suppressed than the diagrams we have been considering; the diagram only
becomes important for small gluon momentum.  This diagram thus provides an
example of how non-linear effects could introduce a natural infrared cutoff
into
our problem.  Unfortunately, it is not easy to resum such diagrams, as is
possible in the two-point function case.

Our result for the 4-point function will take the form
\begin{equation}{\cal O}_{S+P}= \left(\frac{r}{\Lambda}\right)^{2}
R(q^{0})^2 B(p_{1}, \phi_{1}; q^{0}) B(p_{4},
\phi_{4}; q^{0}),\end{equation} where $r$ is a dimensionless constant.  It is
interesting that $r$ may have to be significantly larger than
unity, in order to compensate for the loop-phase-space suppression
factors appearing in diagrams such as Fig.\ref{fig:SD3}.

In place of an explicit infrared cutoff we choose to multiply the kernel of our
equation by the function $k^{2}/(k^{2} + \Lambda^{2}R^{2}(q^{0}))$.
The final equation becomes
\begin{eqnarray}
pB(p, \phi; q^{0}) &=& \frac{3N\alpha}{4\pi^{2}}\int dkd\tilde{\phi}
\frac{kB(k,
\tilde{\phi}; q^{0})k\sin{\tilde{\phi}}}
{\left((q^{0}-k\cos{\tilde{\phi}})^{2}+k^{2}\sin^{2}{\tilde{\phi}}
\right)2\sin{\phi}}
\times \nonumber \\
&& \nonumber \\ &&
\ln{\left(\frac{
(k\sin{\tilde{\phi}}+p\sin{\phi})^{2}+(k\cos{\tilde{\phi}}
-p\cos{\phi})^{2}}
{(k\sin{\tilde{\phi}}-p\sin{\phi})^{2}+(k\cos{\tilde{\phi}}
-p\cos{\phi})^{2}}
\right)}
\times \nonumber \\
&& \nonumber \\
&&\left(1- \frac{q^{0}k\cos{\tilde{\phi}}}{k^{2}}\right)
\times
\frac{k^{2}}
{k^{2}+
\Lambda^{2}R^{2}(q^{0})},
\label{eq:final}
\label{aa}\end{eqnarray}
\noindent where $0 \; ~^{<}_{-} \; \; \phi, \tilde{\phi} \; \;
{}~^{<}_{-} \; \pi$.
We solve for
the function $pB$ in order to increase the accuracy of the
numerical solution that follows, since we expect $B$ to decrease
with increasing $p$.
We note that our
crude representation of the infrared effects does not affect the
functional form of $B$ when $p$ is large.

With regards to the gauge dependence of our results we note that the use
of the Landau gauge, along with a bare massless fermion propagator and a bare
gluon-fermion vertex, is consistent with the Ward-Takahashi
identity at one-loop. Moreover, in the large-$N$ limit only diagrams
$A$ and $B$ are important and the situation is then expected to be
similar to that of the two-point function. In particular in that case
the use of the Landau gauge in the ladder approximation yields results
resembling those found in gauge-invariant treatments \cite{Pen}.

\section{Numerical results}
The form of the integral equation allows us to use the same discretization
lattice for the arguments of the function
$B$ inside and outside the integral.
The angles are discretized according to $\phi(i),\tilde{\phi}(i)
=\frac{i\pi}{(n+1)},
i=1,...,n$. We avoid the endpoints where the angles are equal to
$0$ or $\pi$
since our integral has an integrable singularity there.

The momenta are discretized according to ${\rm log}_{10}p(i) , {\rm
log}_{10}k(i) = {\rm log}_{10}(\Lambda_{IR}+\frac{i-1}{n-1} {\rm
log}_{10}(\Lambda_{UV}/\Lambda_{IR})$, where $\Lambda_{IR, UV}$ are the
infrared and ultraviolet cut-offs.  $\Lambda_{UV}$ does not have to be
much larger than the energy scale where our function is negligibly
small. On the other hand, $\Lambda_{IR}$ has to be smaller than
$\Lambda$ which provides an effective infrared cutoff to our
theory. We choose to work with the ratios $\Lambda/\Lambda_{IR}=10$ and
$\Lambda_{UV}/\Lambda = 100$. We choose the number of points ($n$) in
each dimension to be equal to 40. Our results change little when
this number is increased.

The integral equation is solved via a relaxation method. We first insert
an initial configuration for our function, and then iterate the equation
until it is satisfied to a reasonable accuracy.
We start by setting $q^{0}=0$ and determine the critical coupling necessary for
a solution. We then keep
this coupling fixed, and for different values of $q^{0}$ we compute
$B(p, \phi; q^{0})$ and $R(q^{0})$.
Our critical
coupling is $N \alpha_{c} = 2.7 \pm 0.3$ which, from our previous discussion,
is the same as the one for chiral symmetry breaking via the two-point function
with the same cut-offs.  In the large-$N$ limit and
for infinite $\Lambda_{UV}$ this coupling is given by
$N \alpha_{c} = \frac{2\pi}{3} \approx 2.1$.

\begin{figure}[p]
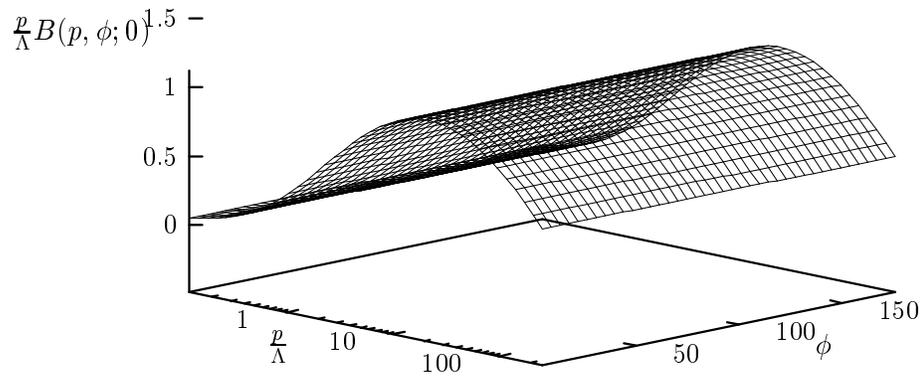

%\vspace{2.cm}
%\input gplot52.tex
\caption{The function
$\frac{p}{\Lambda}B(p, \phi; 0)$}
\label{fig:52}
\end{figure}

The function $B(p, \phi; q^{0})$ has only a weak dependence on $q^0$, and
in Fig. \ref{fig:52} we plot $\frac{p}{\Lambda}B(p, \phi; 0)$.
We see that $\frac{p}{\Lambda}B(p, \phi; 0)$
exhibits the  $\cos{\left(\gamma \log{(p/\Lambda)}\right)}$
behavior (with
$\gamma$ a coupling-dependent constant) that
is well known in two-point function studies.
$B(p, \phi; 0)$ is independent of $\phi$ as expected, and
even for non-zero $q^{0}$ the dependence
on $\phi$ remains very weak.

\begin{figure}[p]
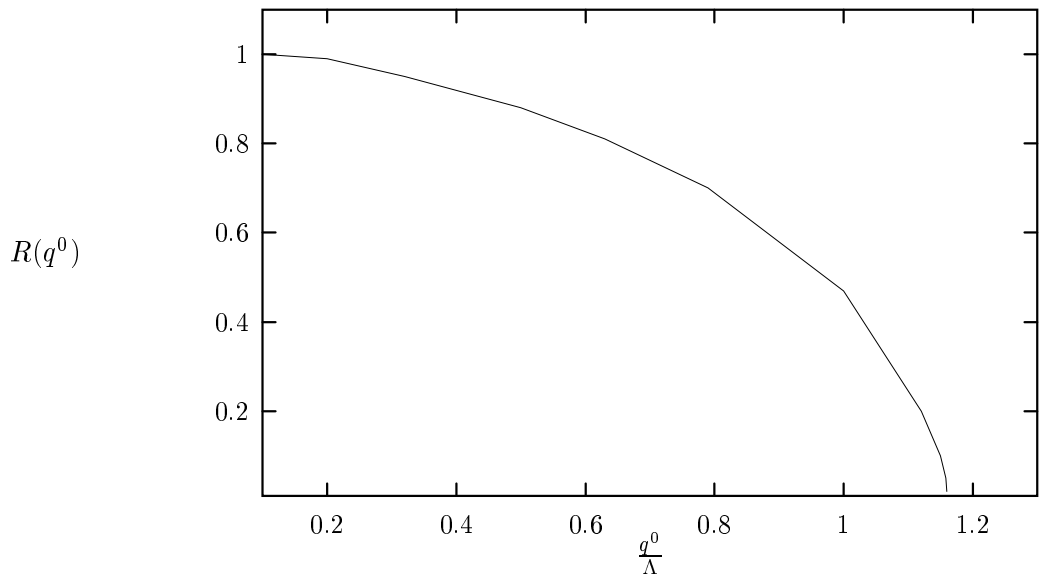

%\vspace{2.cm}
%\input gplot54.tex
\caption{The  function
$R(q^{0})$.}
\label{fig:54}
\end{figure}

The main $q^0$ dependence enters in the function $R(q^{0})$ which we plot
in Fig.\ref{fig:54}.
$R(q^{0})$ falls rapidly with increasing $q^{0}$; we find that
$q^{0}$ cannot exceed $\Lambda$ by much ($\approx 1.16\Lambda$) before
the whole solution collapses to zero.

\section{$2n$-point Green functions}
The form that the integral equation takes for large $N$ suggests a possible
generalization of the above results. In particular,  the diagrams that
dominate are the ones in which a gluon is attached
between two fermion fields
with spinor
indices contracted with each other.
Therefore, one might expect
only condensates of the form
$<\bar{\psi}_{L}\psi_{R}
 ...\bar{\psi}_{L}\psi_{R} + {\rm h. c.}>$, where
 the ellipsis stands for $2n-4$ fermion fields paired with
 each other.  They should form at a critical value of the gauge coupling
similar to that required by the two-  and  four-point functions.

Moreover, we expect the corresponding scalar $2n$-point function $O^{n}_{S+P}$
to factorize as $O^{n}_{S+P}\sim R^n B_{1}...B_{n}$.  Each $B_i$ depends on
the momenta of the respective fermion pair.  In order for each of the $B_{i}$
to satisfy the same final integral equation as before, however, we would have
to assume that the three-momenta of the fermions in each pair are equal and
opposite, so that $\vec{q_{i}}=0$. We expect, though, that our previous
functions $B$ and $R$ should describe the qualitative momentum behavior of
the $2n$-point functions if we are able to go to an optimal reference frame
where the various $\vec{q_{i}}$ are reasonably small.

\section{Conclusions}
In this work we considered chirality-changing fermion four-point functions in
a non-abelian gauge theory which receive non-perturbative contributions
exclusively. We have tackled the problem by writing down an equation derived
from a truncation of the infinite SD hierarchy. The solution to this equation
is  expected to illustrate some generic properties of four-point functions.
The large-$N$ limit of $SU(N)$ renders the four-point problem similar to that
of the  three-point function. We find that a particular four-point function
is much more likely to form than other four-point functions, and that its
critical coupling  is similar to the one required for a two-fermion condensate.
The large-$N$ limit also allows us to discuss the critical and momentum
behavior of fermion chirality-flipping $2n$-point functions for  $n$ higher
than two.
\\\\
\noindent {\Large\bf Acknowledgements} \\
This research was supported in part by the Natural Sciences and
Engineering Research Council of Canada.
\\\\
\noindent {\Large\bf Appendix A}  \\
The purpose of this appendix is to investigate the effects
that the gluons in diagrams $A, ..., F$ have on the color
structure of the fermion vertices. We consider the case of
fermions in a general representation of  a compact simple Lie group.
The group is then defined by the
commutation relations
$[\lambda^{a},\lambda^{b}]= t^{abc}\lambda^{c}$, where $\lambda^{a}$
are the (traceless) generators of the corresponding algebra.

 The Dynkin index $T_{f}$
is defined by the relation $Tr(\lambda^{a} \lambda^{b})=T_{f}\delta^{ab}$.
By  $N_{f} \equiv Tr{\bf 1}$
 we denote the dimension
of the fermion representation,
and by $C_{f}$ and $C_{g}$ the quadratic Casimirs of the
  fermion and adjoint representations.
These are defined by $\lambda^{a} \lambda^{a} = C_{f}{\bf 1}$
and $t^{abc}t^{abc^{\prime}}=
C_{g}\delta^{cc^{\prime}}$. For fermions in the fundamental representation
of $SU(N)$, $N_{f}=C_{g}=N$, $T_{f}=\frac{1}{2}$, and
$C_{f}=\frac{N^{2}-1}{2N}$.

There are two color structures in the problem.
One is {\bf 1}$\otimes{\bf 1}$, which is associated with the functions
${\cal O}_{S+P, T}$,
and the other is $\lambda^{a}\otimes\lambda^{a}$, which is
associated with the functions ${\cal O}_{S+P, T}^{color}$.

\noindent I. VERTEX COLOR STRUCTURE: {\bf 1}$\otimes$ {\bf 1} \\
i. Diagrams $A, B$ \\
We have $\lambda^{a}\lambda^{a}\otimes {\bf 1} = C_{f}{\bf 1}\otimes {\bf 1}
$. Diagrams
 multiplied by $C_{f}$, no change in color structure.

\noindent ii. Diagrams $C, D, E, F$ \\
Color structure  becomes $\lambda^{a} \otimes \lambda^{a}$.

\noindent II. VERTEX COLOR STRUCTURE: $\lambda^{a}\otimes\lambda^{a}$ \\
i. Diagrams $A, B$ \\
$\lambda^{b}\lambda^{a}\lambda^{b}\otimes \lambda^{a}
=(C_{f}-C_{g}/2)\lambda^{a}\otimes \lambda^{a}$.
Diagrams multiplied by $C_{f}-C_{g}/2$, no change in color structure.

\noindent ii. Diagrams $E, F$ \\
Color structure becomes $O_{EF} \equiv
\lambda^{a}\lambda^{b}\otimes \lambda^{a}\lambda^{b}$.
This has to be reduced down to the two initial color structures.

We first expand the matrix $\lambda^{a}\lambda^{b}$ with respect to the
basis of the unity matrix {\bf 1} and the group generators $\lambda^{c}$.
\begin{equation}
\lambda^{a}\lambda^{b} = f^{ab}{\bf 1} + h^{abc}\lambda^{c}
\label{eq:expand}
\end{equation}
By taking the trace of both sides,
we find that $f^{ab}=\frac{T_{f}}{N_{f}}\delta^{ab}$.
Moreover, considering the case $a=b$ proves that $h^{aac}=0$.

The above considerations allow us to write
\begin{equation}
O_{EF}=
\frac{T_{f}}{N_{f}}C_{f}{\bf 1}\otimes {\bf 1} +
h^{abc}h^{abc^{\prime}}\lambda^{c} \otimes \lambda^{c^{\prime}}
\end{equation}

We now have to calculate the quantity $h^{abc}h^{abc^{\prime}}$.
By virtue of (\ref{eq:expand}), we consider the tensor
\begin{eqnarray}
h^{abc}h^{abc^{\prime}}\lambda^{c} \lambda^{c^{\prime}}
& = &
\left(\lambda^{a}\lambda^{b}-\frac{T_{f}}{N_{f}}\delta^{ab}{\bf 1}\right)
\left(\lambda^{a}\lambda^{b}-\frac{T_{f}}{N_{f}}\delta^{ab}{\bf 1}\right)
\nonumber \\
&=&  \left(C_{f}-\frac{C_{g}}{2}-\frac{T_f}{N_{f}}\right)C_{f}
{\bf 1}
\end{eqnarray}
But the unique quadratic tensor proportional to ${\bf 1}$
is the quadratic Casimir of the
fermion representation.
 This means that $h^{abc}h^{abc^{\prime}}=h\delta^{cc^{\prime}}$,
in which
case we find
\begin{equation}
h=C_{f}-\frac{C_{g}}{2}-\frac{T_{f}}{N_{f}}
\end{equation}
So finally we have a new structure
$O_{EF}  =  \frac{T_{f}}{N_{f}}C_{f}{\bf 1}\otimes {\bf 1} +
h\lambda^{c} \otimes \lambda^{c}$, with $h$ given above.
For fermions in the fundamental representation of $SU(N)$, $h=-1/N$.

\noindent iii. Diagrams $C, D$ \\
Color structure becomes $O_{CD} \equiv
\lambda^{a}\lambda^{b}\otimes \lambda^{b}\lambda^{a}$.
By using the commutation relations of the group, we
find that the color structure changes to
$O_{CD}=O_{EF} -\frac{C_{g}}{2}\lambda^{a}\otimes \lambda^{a}$.
\\\\

\pagebreak
%%%%%%%%%%%%%%%%%%%%%%%%%%%%%%%%%%%%%%%%%%%%%%%%%%%%%%%%%%%%%%%%%%%%%%
%%%%%%%%%%%%%%%%%%%%%%%%%%%%%%%%%%%%%%%%%%%%%%%%%%%%%%%%%%%%%%%%%%%%%%
\begin{figure}[p]
%\vspace{7.cm}
%\special{psfile=SDN.ps hoffset=-200 voffset=-200 angle=0
%                       hscale=100  vscale=100}
%\vspace{1.5cm}
\caption{
 An example of a diagram which introduces non-linearities in the infrared.}
\label{fig:SD3}
\end{figure}
%%%%%%%%%%%%%%%%%%%%%%%%%%%%%%%%%%%%%%%%%%%%%%%%%%%%%%%%%%%%%%%%%%%%%%
%
\end{document}